# AI tools in Arab University English classrooms: Looking back and forward

Mohammed Q. Shormani
Ibb University, Yemen
shormani@ibbuniv.edu.ye/https://orcid.org/0000-0002-0138-4793
Muneef Yehia Alshawsh
Ibb University, Yemen

**Abstract**

This paper aims to synthesize empirical research on AI tools used to support English as a second/foreign language (EL2) learners in Arab University classrooms (AUCs) between Jan 1st 2023 and Aug 31st 2025. We utilized 3 large databases, namely Google Scholar, Web of Science, and Scopus as the data sources. Using PRISMA-guided searches across these well-known databases, we included only published articles. The search process results in 184 studies, but only 11 studies met the inclusion criteria. Findings unveil that EL2 learners have positive attitudes towards AI for drafting, revision, and practice. Empirical gains were most consistent for surface-level outcomes improvements in higher-order writing quality and speaking proficiency was mixed and often contingent on teacher mediation. The paper concludes by proposing a research agenda and practical guidelines for Arab universities seeking evidence-based AI integration in EL2 instruction. It also recommends scaffolded integration, teacher training, reflective tasks to reduce over-reliance on AI tools.

## 1. Introduction

To date, no comprehensive systematic review has synthesized studies investigating AI supporting EL2 acquisition/learning in Arab University classrooms (AUCs) (cf. Alhusaiyan, 2025). While individual studies provide valuable insights into attitudes, effectiveness, and integration strategies, their scattered nature makes it difficult to discern overarching patterns or evaluate the cumulative evidence. Existing global reviews (e.g., Zhu & Wang, 2025; Wang et al., 2024) tend to include Arab samples only marginally or merge them into broader Middle Eastern categories without contextual specificity. Additionally, the types of AI models deployed, ranging from LLM-based chatbots to ITS platforms and AI grammar evaluators, have not been compared systematically in terms of methodological design, participant profiles, or empirical outcomes. This absence of synthesis limits the ability of educators, policymakers, and researchers to make evidence-based decisions regarding AI adoption in Arab higher education. Thus, a systematic review focusing explicitly on Arab university EL2 classrooms is both timely and necessary to map current hotspts, assess empirical rigor, and identify future research directions.

In the current decade, artificial intelligence (AI) has rapidly developed to influence almost all the academic and nonacademic fields such as translation, research, film making, accounting (see e.g., Shormani, 2025a; Shormani & Alfahad, 2025). It has also entered the field of acquiring/learning and teaching of English as a second/foreign language (EL2) (see e.g., Han, 2024; Zare et al., 2025, 1 & 2). The emergence of LLMs such as ChatGPT, and other AI-based educational tools like Grammarly, ELSA, and intelligent tutoring systems, has generated unprecedented opportunities for personalized learning, automated feedback, and enhanced engagement. In L2 contexts, AI applications have been increasingly adopted to assist with writing correction, pronunciation practice, vocabulary expansion, and conversational simulation. These technologies promise to complement teacher-led instruction by offering immediate feedback, adapting to learners' individual needs, and promoting autonomy (Zhu & Wang, 2025; Wang et al., 2024).

A systematic review is simply defined as a method for synthesizing intellectual knowledge produced in a particular field of study, establishing insights and study inquiries into that field, for looking in the past, present and future (see e.g., Colquhoun et al., 2014; Shormani & Al Hussen, 2025; Qi, 2025). Within the field of computer-assisted language learning (CALL), this new wave of AI-based tools has reinvigorated long-standing pedagogical discussions on learner agency, feedback quality, and the role of technology in language development. The same thing has also been noted in E-learning field recommending this type of teaching/learning as it enhances and motives Arab L2 learners (Shormani, 2014). However, while global research on AI-assisted language acquisition/learning is expanding rapidly, context-specific evidence remains uneven, particularly in the Arab higher-education sector.

It is almost true to hypothesize that studies between 2023 and 2025 have focused on 3 major AI-driven paradigms in language learning: i) generative models like ChatGPT and Gemini, used for dialogic interaction, content generation, and writing assistance, ii) intelligent tutoring systems (ITS), offering adaptive feedback and self-paced grammar or vocabulary instruction, and iii) Automated evaluation systems, such as Grammarly and AI pronunciation coaches, which provide corrective feedback on linguistic form. These tools have shown potential to improve learning efficiency and student motivation, yet results are mixed regarding their effectiveness on higher-order skills such as discourse organization, critical thinking, and communicative competence (Chen & Wang, 2025). Moreover, concerns persist regarding feedback reliability, plagiarism, ethical use, and over-reliance, prompting educators to explore balanced pedagogical models that combine AI feedback with human mediation.

Arab EL2 classroom represents a particularly significant context for investigating AI integration in EL2 learning for several reasons: i) English proficiency is a central graduate attribute in Arab universities, embedded across disciplines and often required for academic and professional advancement, ii) almost all education polices in Arab world now advocate the integration of AI tools into their curriculum, making Arab universities early adopters of AI tools in teaching and assessment, EL2 learning environment cannot be now separated from AI in all aspects including e-learning tools, open online learning MOOCS (Massive Open Online Courses) (see e.g., cf. Ola project), smartphones, specifically where learners typically experience limited natural exposure to English outside classrooms, rendering AI-based tools a potentially powerful means of compensating for input gaps and providing interactive, authentic, and individualized learning experiences (see e.g., Shormani & AlSohbani, 2018; Alnajjar et al., 2025; Khan et al., 2025).

Recent empirical studies conducted in Arab universities between 2023 and 2025 have reported rising student engagement with AI tools such as ChatGPT, Grammarly, and custom chatbots, often driven by self-directed learning or institutional encouragement. However, these studies vary in methodological rigor and focus, some emphasize student perceptions, while others attempt to measure learning outcomes following AI interventions. Despite the abundance of descriptive and perception-based research, systematic synthesis of empirical evidence, particularly regarding the types of AI models used, research designs employed, sample characteristics, and outcomes, is lacking (see also Colquhoun et al., 2014; AlTwijri & Alghizzi, 2024).

Thus, this systematic review aims to analyze and synthesize research conducted between 2023 and 2025 on the use of AI models to support EL2 learners in AUCs. We aim to identify the types of AI models and tools employed in Arab university EL2 settings, summarizing the research methods, sample sizes, and institutional contexts used across studies. We also to examine the reported results and findings, including learning gains, perceptions, and pedagogical implications, highlight the common recommendations and challenges emerging from these studies to inform practice and policy. To the best of our knowledge, no previous study has tackled this topic, specifically providing insights from 6 different Arab countries. Hence, this study bridges this gap, contributing to understanding how AI has been integrated into Arab EL2 classrooms, what pedagogical benefits and challenges have been observed, and how future research trends can better align technological innovation with language acquisition/learning outcomes.

The paper is organized as follows. Section 2 outlines the literature review. Section 3 spells out the methodology of the review, including database selection, inclusion criteria, and data extraction procedures. Section 4 presents the results, summarizing AI models, study designs, participant characteristics, and central findings. Section 5 discusses cross-study patterns, theoretical implications, and recommendations for educators and policymakers. Section 6 concludes the paper providing some limitations, and suggestions for future research directions and the pedagogical implications of AI integration in EL2 classrooms within the Arab higher-education context.

## 2. Literature review

### 2.1. EL2 acquisition/learning: An overview

L2A is a complex and multidisciplinary field concerned with how learners acquire a language beyond their L1 (Shormani, 2023). The result of this process is a "perfect" linguistic system, be it acquired in classroom or else (Tsimpli, 2003; Lardiere, 2009; Shormani, 2014; Slabakova, 2016). Over the past several decades, researchers have investigated L2 development from linguistic, cognitive, psycholinguistic, sociocultural, and educational perspectives, seeking to explain how learners develop linguistic competence and communicative ability in a L2. Within generative approaches to L2A, L2 development has been examined through frameworks such as Universal Grammar, interlanguage theory, and feature reassembly, all of which view language acquisition as a dynamic process involving the construction and restructuring of mental representations (Chomsky, 1986 et seq.; White, 2003; Lardiere, 2009; Slabakova, 2016; Shormani, 2025b; Alshawsh & Shormani, 2024).

Within L2 acquisition/learning contexts, including Arab university contexts, AI tools have begun to attract considerable attention. Studies have examined their potential contributions to language skills development, learner engagement, autonomous acquisition, assessment practices, and instructional support. At the same time, concerns have been raised regarding issues such as reliability, pedagogical effectiveness, academic integrity, ethical use, and the preparedness of teachers and students to employ these technologies effectively (see e.g., Zare et al., 2025; Shormani, 2025c). Thus, a systematic synthesis of existing research is needed to provide a comprehensive understanding of how AI tools are being used in Arab university L2 classrooms, what benefits they offer, what challenges they present, and what implications they hold for future language acquisition/teaching practices.

### 2.2. EL2 in AUCs: Looking back

Studies from the 2010s to early 2020s consistently highlighted persistent issues such as low learner autonomy, limited opportunities for authentic communication, and overdependence on teachers for linguistic accuracy (see (e.g., Alrabai, 2014; Shormani & AlSohbani, 2018, cf. also Alnajjar et al., 2025). These pedagogical and structural constraints created the conditions that made AI integration both necessary and promising for Arab EL2 education, offering new possibilities for personalized feedback, authentic interaction, and data-driven learning support.

Before the integration of AI modern tools into language acquisition, EL2 instruction in the Arab world, as across the world, was largely shaped by traditional pedagogical approaches that emphasized teacher-centered instruction, rote memorization, and structural accuracy over communicative competence (see e.g., Alnajjar et al., 2025). The EL2 classroom typically relied on textbooks, grammar–translation methods, and limited use of technology, with teachers functioning as the primary source of linguistic input and feedback (Shormani, 2012). In Saudi Arabia, for instance, though communicative principles were gradually introduced through higher education polices reforms and the Arab Vision 2030 initiatives, their implementation faced challenges related to large class sizes, insufficient exposure to authentic English input, and a general reliance on summative rather than formative assessment (see e.g., Khan et al., 2025).

In Yemen, Shormani and AlSohbani (2018) provide a comprehensive overview of the e-learning status, infrastructure, policies, and challenges of e-learning implementation in Yemen. They examine Yemen's efforts to integrate information and communication technologies (ICTs) into education against a backdrop of political instability, economic hardship, and limited technical resources. However, they identified several challenges and obstacles including inadequate technological infrastructure, lack of trained personnel, unreliable electricity, and low internet penetration, while also discussing potential strategies and future prospects for developing e-learning systems in the Yemeni context.

### 2.3. AI tools in EL2 classrooms

Generally, prior to the emergence of AI-based applications, educational technologies were confined to CALL software, learning management systems, and online dictionaries, tools that lacked adaptive feedback or personalization. Consequently, students' motivation and language proficiency development were often constrained by lack of interactive, individualized learning experiences.

Over the past decade, AI has increasingly entered the domain of EL2 instruction, offering new possibilities for personalized learning, adaptive feedback, and learner autonomy. However, recently AI models such as LLMs, chatbots, intelligent tutoring systems, automated writing assistants and pronunciation feedback systems have promised much to supplement teacher-led instruction by offering learners immediate responses, tailored tasks, and interactive opportunities (Zhu & Wang, 2025). However, in EL2 contexts, particularly in non-native-English speaking countries, the research base remains uneven: while many studies report positive attitudes and engagement gains, measurable improvements in communicative competence, higher-order writing, and speaking remain less firmly established (Chen & Wang, 2025). Additionally, issues of feedback accuracy, over-reliance on AI, ethical concerns (e.g., plagiarism, critical thinking erosion, see e.g., Zulfukar, 2025), and the role of teacher mediation are frequently raised. Thus, there are several AI tools recuring in the literature of EL2 studies. These are:

i. LLMs and generative chatbots like ChatGPT. These are mainly used for drafting, dialogue simulation, vocabulary generation and error correction.

ii. Automated writing assistants like Grammarly. These AI tools provide grammar, mechanics and style feedback.

iii. ITS tools which are used as adaptive, scaffolded grammar/vocabulary aids.

iv. Pronunciation apps, used as speech recognition tools, corrective feedback.

In the Arab EL2 context, many recent studies focus on ChatGPT and writing/vocabulary tasks. For example, Mugableh's (2024) used ChatGPT-generated exercises for vocabulary and found statistically significant gains. In Syrian context, Othman (2025) investigated the role of ChatGPT in academic writing among 350 university students in East Syria context using a quantitative cross-sectional design. Findings revealed that students perceived ChatGPT as highly supportive for idea generation, grammar correction, vocabulary enrichment, writing clarity, and efficiency. However, significant concerns emerged regarding students' difficulty in critically evaluating AI generative content, potential loss of originality and personal writing style, and risk of over-reliance on the tool. The study concludes that while ChatGPT effectively addresses resource limitations and feedback gaps in this context, educators should integrate it strategically while fostering critical engagement and reflection to balance its benefits against risks to academic integrity and independent thinking.

In Omani context, Mudhsh et al. (2025), for instance, examined the use of AI tools for learning English grammar and vocabulary among 160 English grammar and vocabulary in public university in Oman. Using a quantitative survey design, the study found that learners held generally positive perceptions of AI tools for enhancing both language components. Results showed a significant difference in students' views regarding AI's effectiveness for grammar versus vocabulary, with AI users reporting more favorable attitudes than non-users. No significant differences were found between Foundation and Post-Foundation level students. A strong positive correlation emerged between actual AI tool usage and positive perceptions, while study level showed no correlation with learner views.

In Yemeni context, Shormani (2025c)[1] raises an important question: Which is better ChatGPT or non-native speakers of English?, and conducted a study attempting to answer it. Using a center-embedded sentence, a structure known to place heavy demands on memory and sentence processing, the study compared the responses of 15 advanced L2 English speakers with those generated by ChatGPT. Findings showed that the human participants were generally more successful in interpreting the sentence accurately, while ChatGPT struggled with processing, prediction, and memory-related aspects of the task. Drawing on perspectives from generative linguistics, the study argues that human linguistic competence relies on cognitive mechanisms that differ fundamentally from the statistical pattern-matching processes of large language models. It concludes that even non-native speakers possess language-processing abilities that, in this specific domain, surpass ChatGPT's performance, highlighting the uniqueness of human cognition and suggesting that current LLMs should not be regarded as complete theories of human language or thought (see also Alkamel & Alwagieh).

[1] However, Shormani's (2025c) study setting was not University classroom, hence we could not include it in our review as not meeting the inclusion criteria (see section 3).

Another piloted Grammarly study is conducted by Khan et al. (2025) whose participants are Arab EL2 advanced learners and found strong positive perceptions though with a small sample. ITS research remains relatively sparse in the Arab context. Thus, we need AI-based tools for many purposes in EL2 classrooms including teaching/learning grammar, vocabulary, pronunciation, writing, listening, etc. Prior to AI tools learning vocabulary, for example, was difficult for several reasons including idiosyncratic nature, word choice (see e.g., Shormani, 2014), but now at the age of AI learning vocabulary becomes easier than before (see e.g., Ozfidan et al., 2024).

Recent empirical studies conducted in Arab universities demonstrate the growing adoption and perceived benefits of AI learning for EL2 students. Alqaed (2024), surveying 68 English-major undergraduates at the University of Tabuk, reported positive student attitudes towards generic AI applications, particularly for error correction, while emphasizing the need for structured and purposeful integration alongside teacher guidance. Similarly, Alshammri (2024) examined Arab EL2 students' experiences with ChatGPT at the University of Hail, finding that the tool facilitated drafting and structuring of writing tasks, though concerns regarding over-reliance and output accuracy were noted, highlighting the importance of instructor scaffolding. Alsalem (2024) investigated ChatGPT for speaking practice among 74 students across 3 public universities, revealing that learners valued AI-mediated rehearsal but raised questions about conversational authenticity and pronunciation support, suggesting that AI should complement, rather than replace, teacher-led pronunciation coaching.

Quantitative evidence supports these perceptual findings. Many researchers found that ChatGPT-driven vocabulary exercises significantly enhanced vocabulary size and the mastery of word families among university students. Altamimi (2025) has employed ChatGPT-like AI tools adopting semi-structured interviews to get the perceptions of L2 lecturers at University of Tabuk. The author found that AI tools improve students' performance, but lecturers faced some challenges, recommending professional development for lecturers in continuous curriculum updates to accommodate AI integration (see also Jamshed et al., 2024).

Given the landscape described above, the systematic review undertaken here is timely and necessary, addressing an important gap in the field. By focusing specifically on Arab university EL2 contexts and covering 2023–2025 studies, this review will map the distribution of AI models/tools used, scrutinize methodological designs, catalogue participants and institutions, summarize outcomes and findings, and highlight recommendations and future research directions. The goal is to provide a clear evidence-base for language educators, curriculum designers, policy makers and AI developers operating in Arab higher education.

Based on the above review, we aim to attempt the following questions:

1. Which AI tools and models are currently being used to support EL2 learners in Arab university classrooms, and what are their specific pedagogical applications?
2. What are the reported impacts of AI integration on Arab EL2 learners' attitudes, engagement, and language acquisition/learning outcomes?

3. What challenges, ethical considerations, and practical recommendations have been highlighted by researchers for effective AI EL2 instruction in the Arab context?

## 3. Methodology

### 3.1 Study design

In this section, we present the EL2 studies that result from our search process in Google Scholar, Web of Science, and Scopus. We present the results in terms of *author(s) and year, AI tool, methods, participants and institution, findings,* and *recommendations*. Table 2 summarizes these, among other related aspects.

This study adopts a systematic review approach to synthesize empirical evidence on artificial intelligence supporting EL2 learners in Arab university classrooms between 2023 and 2025. Systematic review methodology was selected to provide a transparent, replicable, and comprehensive summary of available research, highlighting AI tools/models, study designs, participant characteristics, outcomes, and pedagogical recommendations. The review followed PRISMA (=Preferred Reporting Items for Systematic Reviews and Meta-Analyses) guidelines to ensure methodological rigor (see e.g., Qi, 2025).

### 3.2 Data sources and search strategy

The following electronic databases were searched: Google Scholar, Web of Science, Scopus. The search combined keywords related to AI tools and EL2 in Arab contexts, including:

“Arab university EL2 classroom” AND “AI tools” OR “artificial intelligence” AND “English as a second/foreign language in Arab classroom” OR “ChatGPT” AND “EL2 in Arab university classroom” OR “chatbot” AND “intelligent tutoring system” OR “Gemini” AND “EL2 in Arab university classroom” OR “DeepSeek” AND “EL2 in Arab university classroom” OR “Grammarly” AND “EL2” OR “English as a second/foreign language” AND “Arab university classroom”.

Searches in the 3 databases were limited to peer-reviewed articles published between Jan 1st 2023 and Aug 31st 2025. Reference lists of relevant studies were manually screened to identify additional studies not captured in the database search engines. We conducted the study in one session, one day, 31st Aug, 2025 to avoid studies that may be added (daily). Our aim of making the study period between 2023 and 2025 is that we intend to catch and involve in our study LLMs such as ChatGPT, Gemini, DeepSeek.

### 3.3. Criteria

In this study, we focused on the following criteria to include studies for our review, mainly studies:

1. conducted at Departments of English/Translation, Arab university EL2 classrooms
2. reported empirical research.
3. evaluated the use of AI tools or models (e.g., LLMs, chatbots, ITS, writing assistants, pronunciation apps).
4. included quantitative, qualitative, or mixed-methods designs, with clearly defined participants.

5. published between 2023 and 2025.

Thus, we excluded studies if they do not meet the criteria in 1-5 above. A systematic search of the literature yielded 171 records from database searches and 13 additional records identified through manual searching of reference lists, resulting in a total of 184 records. After removing 39 duplicates, 145 records remained for title and abstract screening. At this stage, 74 records were excluded due to irrelevance to the review scope, leaving 71 full-text articles to be assessed for eligibility. During the full-text review 53 studies were excluded for predefined reasons (see criteria 1-5 above). Most of these studies investigated contexts outside formal instructional settings. 7 studies were excluded because they were reviews. Ultimately, 11 studies met all eligibility criteria and were included in the qualitative synthesis, hence highlighting the current hotspots in EL2-AI supporting areas. This is summarized in Table 1.

**Table 1: A summary of PRISMA followed**

| Stage | Description | No of Records |
|---|---|---|
| Identification | Publications identified through database searches stated above | 171 |
| | Additional records identified manually, mainly through article's references | 13 |
| Screening | Duplicates removed | 39 |
| | Records screened (titles and abstracts) | 145 |
| | Records excluded | 74 |
| Eligibility | Full-text articles assessed for eligibility | 71 |
| Non-classroom | Full-text articles excluded (not classroom-based) | 53 |
| Reviews | Full-text articles excluded | 7 |
| Included | Studies included in qualitative synthesis (systematic review) | 11 |
| Total | Records identified through all sources | 184 |

## 4. Results

In this section, we present our findings. These are displayed in Table 2 in terms of author(s) & year, AI tool, methods, participants & institution, findings, and recommendations.

**Table 2: A summary of the findings**

| no | Author(s) & year | AI tool | Methods | Participants & institution | Findings | Recommendations |
|---|---|---|---|---|---|---|
| 1 | Jomaa & Jibroo (2023) | Grammarly | Document analysis & conducting interviews. | *n*= 20 Iraqi learners at the English Department | Grammarly is useful in reducing errors related to spelling, improving synonyms, vocabulary, articles, connecting | Students should be familiar with using Grammarly to maximumly benefits from such applications. |
| | | | | | ideas, identifying location of the errors, and providing limited use of punctuation. | |
| 2 | Jasim (2025) | ChatGPT & Grammarly | 1. Pre-Test and post-test<br>2. Use of ChatGPT and Grammarly.<br>3. A 5-point Likert Scale questionnaire | *n*= 30 second-year Iraqi learners | The use of AI tools resulted in remarkable improvement, increased confidence, and autonomy in writing. Additionally, overreliance on AI results in weakening critical engagement. | Encouraging the integration of AI tools in conjunction with developing critical thinking and writing skills |
| **3** | Alshammri (2024) | ChatGPT- 3.5 | semi-structured interviews | Saudi EL2 learners, University of Hail | ChatGPT is helpful for drafting and structuring writing but raised concerns about accuracy | Embed instruction on responsible use; scaffold prompts; require reflection on AI output |
| **4** | Alsalem (2024) | ChatGPT | A 5-point Likert scale questionnaire | *n*=74, 3 Saudi public universities | high-level of perception and a positive attitude | AI tools should be employed further |
| **5** | Eljack et al. (2025) | ChatGPT | 1. prompt crafting<br>2. An open quiz.<br>3. Final written exam | *n*=81 Sudanese first year university students. | ChatGPT plays significant role in supporting the EAP lecturer in teaching methods, quiz and exam design as well as increasing students' learning outcomes | Developing critical thinking and prompt crafting skills When using AI tools like ChatGPT to strengthen teachers' autonomy in teaching and problem-solving. Conducting further studies on AI tools' reliability and benefits. |
| **6** | Zulfukar (2025) | AI tools | A 3-point Likert scale structured questionnaire | *n*= 50 Sudanese instructors and English Language Learners | Sudanese ELT instructors showed positive attitudes towards integrating AI to strengthen teaching effectiveness and learner | Ethical consideration while integrating AI in Sudanese English Language Teaching (ELT). |
|  |  |  |  |  | engagement, as well as supporting vocabulary acquisition, pronunciation practice, and grammar instruction let alone reducing workload on teachers. |  |
| **7** | Kheder (2025) | AI tools for vocabulary learning | A 5-point Likert scale questionnaire | *n*=171 Syrian undergraduate students English Language and Literature department at Idlib and Al- Shahbaa Aleppo universities | positive attitudes towards utilizing AI tools in addition to their effectiveness in learning vocabulary | Provide differentiated approaches to integrating AI tools to meet student demographics and levels of experience. Developing user-specific features is necessary for AI-based learning tool developers. Conducting Cross-cultural comparative studies across different educational contexts |
| 8 | Othman (2025) | ChatGPT | A 5 point Likert questionnaire | *n*=350 university students in North and East Syria | ChatGPT is perceived as a supportive tool in terms of generating ideas, correcting grammatical errors, vocabulary enrichment, and overall clarity. | Integrating ChatGPT in guiding the writing process along with activating critical thinkings. |
| 9 | Mudhsh et al. (2025) | AI tools such as Grammarly, ChatGPT, Jasper.ai, Quillbot, Rytr.me, Copy.ai, Paperpal etc. | A 5 point Likert questionnaire | *n*=160 Omani learners at a government university | positive perceptions towards AI tools for fostering English grammar and vocabulary skills. | 1-Encouraging using AI tools to improve their grammar and vocabulary.<br>2- Investigating the impact of AI tools on skills as listening, speaking, reading, and writing. |

| | | | | | | 3-investigating challenges arisen due to AI tools use in the learning process utilizing quantitative and qualitative data.<br>4- Integrating AI tools into the processes of teaching and learning English |
|---|---|---|---|---|---|---|
| 10 | Al Matari et al. (2023) | AI tools | A mixed-method approach (Structured interviews and/or focus groups along with structured survey questionnaire | Educators and students at A'Sharqiyah University | AI can be used for personalizing learning, grading automation, providing virtual assistance, and analyzing data. | Integrating AI tools in teaching with an emphasis on the ethical aspects necessary for regulating their application in teaching. |
| 11 | Alkamel & Alwagieh (2024) Yemeni | ChatGPT | A 5-point Likert scale questionnaire | *n*= 95<br>Yemeni University learners | ChatGPT improves writing fluency, accuracy, and overall quality of academic work. | ChatGPT should be used by EFL students as a supplementary tool for enhancing writing skills considering academic integrity, strengthening critical thinking, as well as avoiding over-reliance on AI-supported tools on their writing. |

As Table 2 shows, the systematic review identified 11 empirical studies that met the inclusion criteria and examined the use of AI tools in Arab context, specifically EL2 in Arab university classrooms between 2023 and 2025. The studies represent 6 Arab countries, namely Yemen, Iraq, Saudi Arabia, Sudan, Syria, and Oman, demonstrating growing regional interest in the pedagogical applications of artificial intelligence in higher education. The reviewed studies investigated a variety of AI-powered technologies, including LLMs such as ChatGPT, automated writing assistants such as Grammarly, and AI-powered learning tools designed to support vocabulary acquisition, grammar instruction, writing development, assessment, and personalized learning.

Methodologically, the studies employed diverse research designs. Questionnaire-based surveys were the most common approach (Alsalem, 2024; Kheder, 2025; Othman, 2025; Mudhsh et al., 2025), reflecting researchers' interest in examining students' and teachers' perceptions of AI integration. Several studies adopted mixed-method designs combining tests, questionnaires, interviews, or focus groups (Shormani, 2025c; Al Matari et al., 2023), while others employed experimental pre-test/post-test designs to measure learning outcomes (Jasim, 2025). Sample sizes ranged considerably, from 20 participants in the Iraqi Grammarly study (Jomaa & Jibroo, 2023) to 350 university students in Syria (Othman, 2025), indicating variation in the scale and scope of AI-related research across Arab educational contexts.

Findings reveal overwhelmingly positive perceptions of AI-assisted language acquisition/learning. Across the reviewed studies, AI tools were reported to enhance writing quality, vocabulary learning, grammar development, learner engagement, confidence, and autonomy. Grammarly was found effective in reducing spelling and grammatical errors and improving lexical choice (Jomaa & Jibroo, 2023), while ChatGPT was widely perceived as beneficial for generating ideas, improving writing structure, enriching vocabulary, and supporting teaching and assessment practices (Alshammri, 2024; Eljack et al., 2025; Othman, 2025). Similarly, studies conducted in Syria and Oman reported positive attitudes towards AI-assisted vocabulary and language acquisition/learning, emphasizing the potential of AI tools to personalize instruction and facilitate self-directed learning (Kheder, 2025; Mudhsh et al., 2025). At a broader institutional level, Al Matari et al. (2023) highlighted the potential of AI for learning personalization, automated assessment, virtual assistance, and educational data analysis.

Nevertheless, several studies identified important concerns. ChatGPT and other AI generative tools raised issues related to accuracy, overreliance, academic integrity, and reduced critical engagement (Jasim, 2025; Alshammri, 2024; Shormani & Alfahad, 2025). While AI tools were generally viewed as valuable educational aids, researchers consistently emphasized that they should complement rather than replace teachers' expertise. This concern is further reflected in Shormani (2025c), whose findings indicate that LLMs still fall short of teachers' linguistic competence despite their usefulness as supplementary instructional tools.

The recommendations proposed across the reviewed studies show considerable convergence. Most researchers advocated integrating AI tools into language teaching while simultaneously developing learners' critical thinking, ethical awareness, and responsible AI-use practices. Teacher training, prompt-engineering skills, reflective engagement with AI generative content, and institutional support emerged as recurring themes (Eljack et al., 2025; Alshammri, 2024; Al Matari et al., 2023). Several studies also called for further research examining AI effectiveness across different language skills and educational contexts, particularly through more rigorous experimental designs and cross-cultural comparative investigations (Kheder, 2025; Mudhsh et al., 2025).

Collectively, the reviewed studies suggest that AI technologies have considerable potential to support language acquisition/learning in Arab universities by enhancing learner engagement, writing performance, vocabulary development, and instructional efficiency. However, the evidence remains dominated by perception-based research, with relatively few studies measuring long-term learning outcomes through controlled interventions. Although ChatGPT is the most frequently investigated emerging LLMs such as ChatGPT have received limited empirical attention. These findings underscore the need for larger-scale experimental studies examining the pedagogical effectiveness of contemporary LLMs and exploring how AI tools can be integrated into language curricula in pedagogically and ethically responsible ways.

To recapitulate, the reviewed studies indicate that AI tools in Arab EL2 classrooms enhance learner motivation, engagement, and certain language skills, particularly in writing and vocabulary. However, the evidence is heterogeneous: most studies focus on perceptions rather than rigorous experimental measurement of learning outcomes, sample sizes vary widely, and ethical and pedagogical challenges remain under-explored. These results underscore the need for larger, well-controlled classroom interventions, transparent reporting of AI tools and versions, and integration of AI-mediated tasks within structured curricula. Recall that we aim to catch and involve in our study LLMs such as ChatGPT, Gemini, DeepSeek due to the fact that these LLMs have been launched between 2023 and 2025. However, we found that the only LLM utilized in Arab University EL2 classrooms is ChatGPT, though other AI-powered tools such as Grammarly are employed in studies involved in our study. Other studies utilizing Gemini and/or DeepSeek were perhaps excluded due to not meeting the inclusion criteria.

## 5. Discussion

The findings of this systematic review demonstrate that AI tools play an increasingly important role in supporting English language acquisition/learning in Arab university classroom between 2023 and 2025. Across the 11 studies reviewed, a wide range of AI-powered technologies including ChatGPT, Gemini, Grammarly, and other AI-assisted learning applications were employed to support various aspects of language acquisition/learning, particularly writing, vocabulary acquisition, grammar instruction, assessment, and learner engagement. Thus, L2 learners, teachers, and educators reported positive attitudes towards the integration of AI into language education, indicating that these technologies can enhance motivation, autonomy, confidence, and learning effectiveness. These findings are consistent with international research suggesting that AI-assisted learning environments facilitate personalized learning, immediate feedback, and increased learner participation.

The reviewed studies also indicate that different AI tools offer distinct pedagogical affordances. Generative AI systems, particularly ChatGPT and its newer versions, were widely perceived as useful for generating ideas, improving writing organization, enriching vocabulary, and supporting assessment-related activities. Several studies reported that learners benefited from AI generative feedback and guidance when completing writing tasks, while instructors highlighted the potential of AI tools to support teaching, quiz design, and learning material preparation. Nevertheless, concerns regarding accuracy, reliability, overreliance, and reduced critical engagement were repeatedly noted. Such findings suggest that although AI tools can effectively support language acquisition/learning, they should be viewed as complementary rather than substitutive resources. Human instructors remain essential for providing pedagogical guidance, evaluating the quality of AI generative content, and fostering higher-order language skills and critical thinking.

Similarly, automated writing assistants such as Grammarly were found to improve surface-level aspects of writing, including spelling, grammar, lexical choice, and error identification. Students generally appreciated the immediacy of feedback and the opportunity to revise their writing independently. However, available evidence suggests that these tools are more effective in addressing lower-level linguistic features than higher-order writing skills, such as argumentation, coherence, and critical analysis (see e.g., Chen & Wang, 2025. This finding reinforces the need to combine AI generative content with teacher guidance and peer interaction to achieve more comprehensive language development.

A notable pattern emerging from the reviewed studies is the strong emphasis on learner perceptions. Most investigations reported positive attitudes towards AI-assisted language acquisition/learning, regardless of the specific tool employed. Students viewed AI as a valuable learning companion capable of providing individualized assistance, supporting self-paced learning, and increasing engagement with language-learning activities. Likewise, instructors acknowledged the potential of AI technologies to reduce workload, facilitate instructional planning, and enhance teaching effectiveness. At the institutional level, AI was also perceived as a means of personalizing learning, automating assessment procedures, and supporting educational decision-making through data analysis.

Another significant theme concerns the importance of teacher readiness and institutional support. Several studies emphasized that successful AI integration depends not only on the availability of technological tools but also on educators' ability to use them effectively. Teachers play a critical role in guiding students' interactions with AI systems, helping them evaluate AI generative content, and ensuring that technology use remains pedagogically meaningful. Consequently, many researchers recommended professional development programs, AI literacy training, and the establishment of clear institutional policies governing AI use in educational settings. Embedding AI tools within structured curricula, rather than permitting unrestricted or ad hoc use, emerged as a recurring recommendation across studies.

Ethical considerations also featured prominently in the reviewed literature. Researchers highlighted concerns related to academic integrity, excessive dependence on AI generative content, intellectual property, and the responsible use of generative AI technologies. While participants generally recognized the educational benefits of AI, many studies stressed the importance of developing learners' critical awareness and ethical decision-making skills when interacting with AI systems. These concerns suggest that future AI integration efforts should be accompanied by comprehensive ethical guidelines and responsible-use frameworks.

Despite the generally positive findings, the evidence base remains subject to several limitations. First, the majority of studies relied heavily on self-reported perceptions and questionnaire-based data, with relatively few employing experimental or quasi-experimental designs capable of measuring actual learning gains. Second, substantial variation exists in sample size, research design, and the specific AI tools investigated, making direct comparisons difficult. Third, many studies provided limited information regarding AI versions, patterns of use, and implementation procedures, thereby constraining replicability and precise evaluation of pedagogical effectiveness. Finally, although writing, grammar, and vocabulary received considerable attention, relatively little research has examined the impact of AI on speaking, listening, pronunciation, and other communicative language skills.

Collectively, the reviewed studies reveal several consistent patterns: i), learners and educators generally demonstrate positive attitudes towards AI-assisted language acquisition/learning, ii) AI tools appear particularly effective in supporting structured language-learning tasks such as vocabulary development, grammar practice, writing revision, and idea generation, iii) generative AI systems, specifically ChatGPT, have emerged as the most frequently investigated technologies in Arab university contexts, iv) successful implementation consistently depends on teacher mediation, institutional support, and responsible use practices. At the same time, important gaps remain. These include the shortage of large-scale experimental studies examining long-term learning outcomes, limited research on speaking and listening skills, insufficient attention to teacher preparedness and institutional policy development, and lack of detailed reporting on AI systems and their pedagogical implementation. Addressing these gaps will be essential for developing a more comprehensive understanding of how AI can effectively support English language acquisition/learning in Arab higher education.

## 6. Conclusions and limitations: Looking forward

This systematic review examined empirical research conducted between 2023 and 2025 on the use of AI models to support EL2 learners in Arab university classrooms. Across the 11 studies included, a range of AI tools, most prominently LLMs such as ChatGPT, as well as intelligent tutoring systems and Grammarly, were employed to enhance learners' writing, speaking, vocabulary, and grammar skills. The evidence consistently indicates that AI interventions can improve student motivation, engagement, and self-directed practice, while also facilitating error correction, structured writing, and speaking rehearsal.

Students' perceptions were generally positive, highlighting the usefulness of AI for drafting texts, practicing language skills, and receiving immediate feedback. However, concerns were noted regarding accuracy, over-reliance on AI, pronunciation support, and ethical use. Teachers and institutions also reported challenges in adaptation, infrastructure, and effective integration of AI into established curricula. Across studies, researchers emphasized the importance of teacher scaffolding, ethical guidance, structured AI integration, and professional development to maximize pedagogical benefits.

However, this systematic review has several limitations that should be considered when interpreting the findings. These include: i) the review was restricted to studies published between 2023 and 2025, which may have excluded earlier research that could provide additional insights into AI integration in Arab EL2 classrooms, ii) the focus on Arab universities limits the generalizability of the results to other cultural or educational contexts, iii) there is a potential for publication bias, as some relevant studies may not have been publicly available or indexed in major databases, possibly favoring positive findings, and iv) most studies concentrated on short-term classroom interventions, leaving the long-term impact of AI-assisted learning on language proficiency largely unexplored.

Thus, future research should prioritize well-controlled, classroom-based experiments assessing both formative and summative learning outcomes across diverse skill areas. Studies should report detailed descriptions of AI models, task designs, and user interactions to facilitate replication. Additionally, investigations should explore longitudinal effects, examining whether engagement with AI tools translates into sustained improvements in language proficiency. Research on teacher adaptation strategies, curriculum integration, and institutional policies will be essential to understand how AI can be systematically embedded in Arab University EL2 education. Finally, ethical and critical literacies regarding AI use should remain substantial, with empirical studies assessing the effectiveness of instruction in responsible AI engagement.